\documentclass[sigconf]{acmart}

\usepackage[most]{tcolorbox}
\usepackage{algpseudocode}
\usepackage{microtype}
\usepackage[ruled,vlined,linesnumbered]{algorithm2e}
\usepackage{amsmath}  
\usepackage{enumitem}

\AtBeginDocument{%
  }

\copyrightyear{2026}
\acmYear{2026}
\setcopyright{cc}
\setcctype{by}
\acmConference[WWW '26]{Proceedings of the ACM Web Conference 2026}{April 13--17, 2026}{Dubai, United Arab Emirates}
\acmBooktitle{Proceedings of the ACM Web Conference 2026 (WWW '26), April 13--17, 2026, Dubai, United Arab Emirates}
\acmPrice{}
\acmDOI{10.1145/3774904.3792070}
\acmISBN{979-8-4007-2307-0/2026/04}

\begin{document}

\title{ThinkRec: Thinking-based Recommendation via LLM}
\author{Qihang Yu}
\authornote{Equal contribution.}
\orcid{0009-0000-2650-5080}
\affiliation{
  \institution{Zhejiang University}
  \city{Hangzhou}
  \country{China}
}
\email{yuqihang@zju.edu.cn}

\author{Kairui Fu}
\authornotemark[1]
\orcid{0009-0004-0284-671X}
\affiliation{
  \institution{Zhejiang University}
  \city{Hangzhou}
  \country{China}
}
\email{fukairui.fkr@zju.edu.cn}

\author{Zheqi Lv}
\authornotemark[2]
\orcid{0000-0001-6529-8088}
\affiliation{
  \institution{Zhejiang University}
  \city{Hangzhou}
  \country{China}
}
\email{zheqilv@zju.edu.cn}

\author{Shengyu Zhang}
\authornotemark[2]
\orcid{0000-0002-0030-8289}
\affiliation{
  \institution{Zhejiang University}
  \city{Hangzhou}
  \country{China}
}
\email{sy_zhang@zju.edu.cn}

\author{Xinhui Wu}
\orcid{0009-0000-1533-0545}
\affiliation{
  \institution{Ant Group}
  \city{Hangzhou}
  \country{China}
}
\email{18155120673@163.com}

\author{Chen Lin}
\orcid{0009-0005-1571-2259}
\affiliation{
  \institution{Ant Group}
  \city{Hangzhou}
  \country{China}
}
\email{chenlin9@connect.hku.hk}

\author{Feng Wei}
\orcid{0009-0005-6928-1685}
\affiliation{
  \institution{Ant Group}
  \city{Shanghai}
  \country{China}
}
\email{mefwei@gmail.com}

\author{Bo Zheng}
\orcid{0009-0001-4007-3955}
\affiliation{
  \institution{Ant Group}
  \city{Hangzhou}
  \country{China}
}
\email{zhengbo_321@163.com}

\author{Fei Wu}
\authornote{Corresponding author.}
\orcid{0000-0003-2139-8807}
\affiliation{
  \institution{Zhejiang University}
  \city{Hangzhou}
  \country{China}
}
\affiliation{
  \institution{Shanghai AI Laboratory}
  \city{Shanghai}
  \country{China}
}
\email{wufei@zju.edu.cn}

\renewcommand{\shortauthors}{Qihang Yu et al.}
\begin{abstract}
Recent advances in large language models (LLMs) have enabled more semantic-aware recommendations through natural language generation. Existing LLM for recommendation (LLM4Rec) methods mostly operate in a System 1-like manner, relying on superficial features to match similar items based on click history, rather than reasoning through deeper behavioral logic.
This often leads to superficial and erroneous recommendations.
Inspired by this, we propose ThinkRec, a thinking-based framework that shifts LLM4Rec from an intuitive system to a rational system. First, ThinkRec introduces a thinking activation mechanism by injecting synthetic reasoning traces, making the recommendation process resemble the Chain of Thought (CoT) reasoning of LLMs. This mechanism analyzes interaction histories, identifies user preferences, and makes decisions based on target items.
Furthermore, considering the highly diverse distribution of recommendation data, we propose an instance-wise expert fusion mechanism to reduce the reasoning difficulty.
By dynamically assigning weights to expert models based on users' latent features, ThinkRec adapts its reasoning path to individual users, thereby enhancing precision and personalization.
Extensive experiments on various real-world web user behavior preference datasets demonstrate that ThinkRec significantly outperforms baselines in terms of recommendation accuracy and interpretability, providing superior recommendations based on a deeper understanding of user intent and a more rigorous reasoning process. Code is available in \url{https://github.com/Yu-Qi-hang/ThinkRec}.

\end{abstract}

\ccsdesc[300]{Information systems}
\ccsdesc[500]{Information systems~Information retrieval}
\ccsdesc[300]{Information systems~Retrieval tasks and goals}
\ccsdesc[500]{Information systems~Recommender systems}

\keywords{LLM, Recommendation System, Reasoning Model}


\maketitle

\section{Introduction}
Recommendation systems are indispensable in modern digital platforms, enabling users to navigate vast content efficiently \cite{glenski2021improving,ma2023punr,iana2024train}. 
Traditional sequential recommendation methods rely on implicit modeling of user interaction histories and are limited in their ability to model context and incorporate broader knowledge, which restricts their reasoning and generalization capabilities.
Recent advances in LLMs offer strong contextual comprehension and extensive world knowledge to improve recommendation systems.

\begin{figure*}[tb]
  \centering
  \includegraphics[width=0.95\textwidth]{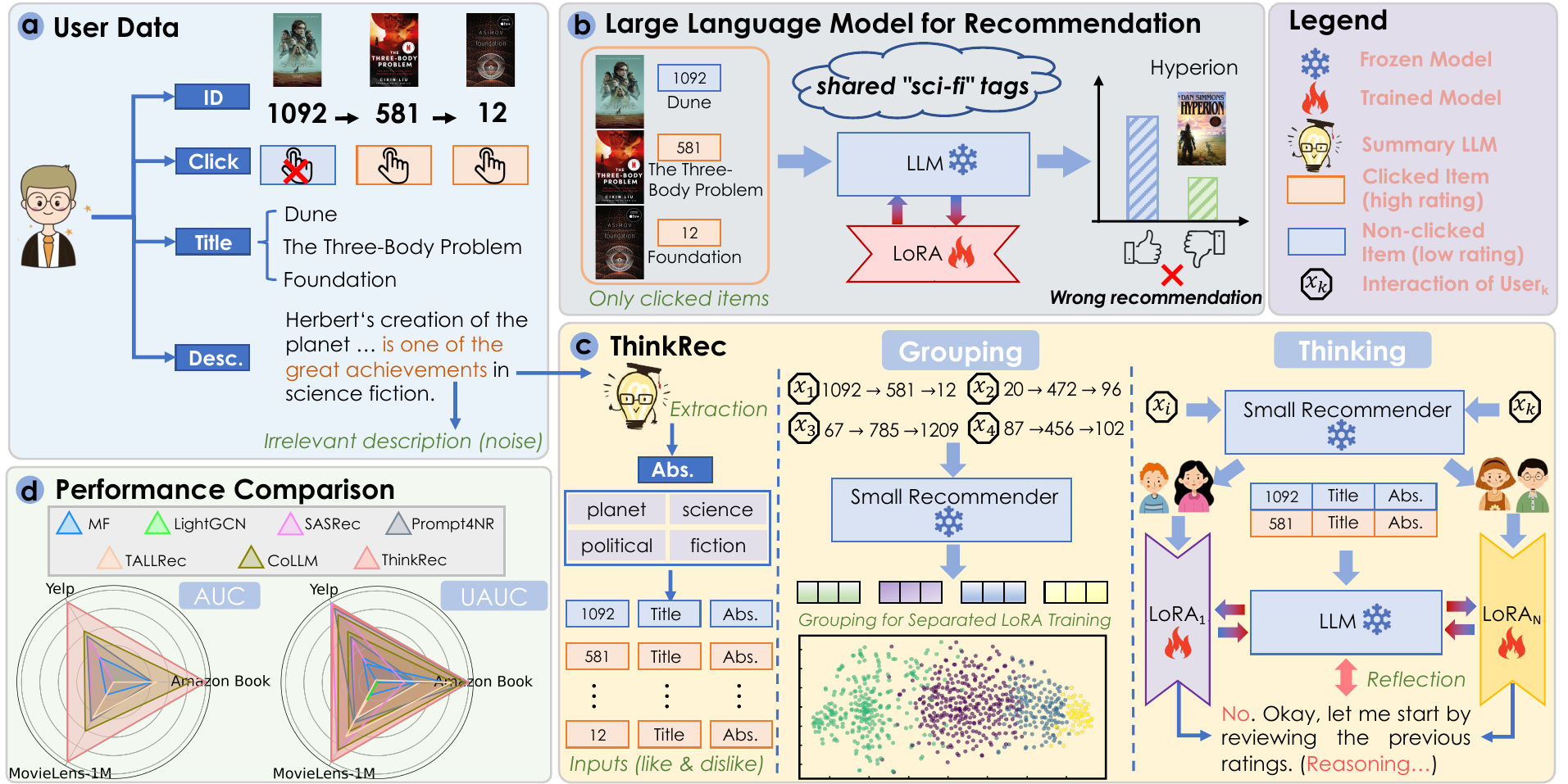}
    \vspace{-0.3cm}
  \caption{(a) shows the composition of user interaction data. (b) and (c) illustrate previous LLM-based recommendations and our ThinkRec, respectively. (d) compares ThinkRec with baselines in three real-world datasets.}
\Description{(a) shows the composition of user interaction data. (b) and (c) illustrate previous LLM-based recommendations and our ThinkRec, respectively. (d) compares ThinkRec with baselines in three real-world datasets.}
  \label{fig:intro}
  \vspace{-0.35cm}
\end{figure*}
Prior LLM4Rec paradigms can be separated into three categories: (1) \textit{item scoring} \cite{liu2024sequential,zhang2025collm}, where LLMs answer binary preference questions given user and item context; (2) \textit{item generation} \cite{li2023prompt,bao2024decoding,bao2025bigrec,fu2025forge}, which maps natural language prompts to item IDs through aligned embedding spaces or semantic identifiers; and (3) \textit{hybrid modeling} \cite{geng2023vip5,zhang2024agentcf}, where a single LLM performs multiple tasks such as point-wise prediction, pair-wise ranking, or list-wise generation.
Although these LLM4Rec methods differ in output formats and representation learning strategies, they fundamentally resemble \textit{System 1}—the intuitive system—in cognitive science~\cite{kahneman2011thinking}. They tend to match similar items based on click history, while overlooking the deeper behavioral logic.
This limitation becomes evident in cases such as the one shown in Figure~\ref{fig:intro}(b), where the user’s behavior over time is: dislikes ``Dune'', likes ``The Three-Body Problem'', and likes ``Foundation'' (all three are science fiction). Methods that rely on System 1-like intuition tend to infer that the user would also like ``Hyperion'' (science fiction) simply because it belongs to the same genre. In reality, the user might dislike philosophical or metaphysical themes in fiction, which are prominent in *Hyperion*, making it an unsuitable recommendation. Clearly, if we can leverage the vast knowledge in LLMs and fully activate their reasoning capabilities
, recommendation performance can be significantly improved.

\textit{This motivates our effort to push LLM4Rec from a System 1 paradigm toward a more rational, System 2-like reasoning framework}. Recent concurrent works have also explored latent reasoning \cite{tang2025thinkrecommendunleashinglatent,zhang2025reinforcedlatentreasoningllmbased} or dual-head reasoning architectures \cite{you2025r2eclargerecommendermodels} to address this.
We raise two key questions for the advancement:
1) How to effectively balance recommendation objectives with language modeling tasks to fully exploit the reasoning capabilities of LLMs. Existing methods often prioritize direct recommendation metrics such as hit rate or ranking accuracy, overlooking LLMs' strengths in semantic reasoning. However, blind reinforcement thinking can lead to simple next token prediction, defeating the goal of recommendation. 2) How to think more effectively in the presence of diverse user behaviors and underlying preferences. As shown in Figure~\ref{fig:intro}(c), user behaviors vary widely, and uniform modeling tends to obscure personality preferences while distracting the model from salient signals, impairing its reasoning capacity. Moreover, inferring user intent solely from high-rating items and generic world knowledge limits the informational basis for accurate preference reasoning.

To address these challenges, we propose \textbf{\underline{Think}}ing-based \textbf{\underline{Rec}}-ommendation via LLM, abbreviated as \textbf{ThinkRec}. One of the main problems faced is that the data and optimization goals of recommendation tasks lack the ability to activate thinking in LLMs (Challenge 1). To overcome this challenge, we designed the item augmentation and thinking activation framework for finetuning. The fine-tuned model analyzes associations in historical item information, determines user preferences, and gives explicit reasons while deriving recommendations. We extracted the metadata keywords of the items with the help of an existing summarization model as the augmentation information of the items to support the reasoning. In addition, reasoning data is synthesized using a strong reasoning model, and the reasoning capability is distilled to the local model by mixed sampling of reasoning and recommendation data. Therefore, item augmentation and thinking activation become a bridge connecting the recommendation task and the language task, making recommendations traceable. To address the difficulty of thinking diversely with rich information (Challenge 2), we add the user's preferences (yes/no) of items to the prompts and generate personalized recommendation experts based on latent user features. In the technique, we design a dynamic Low-Rank Adaptation (LoRA) fusion method. Users are grouped by latent user features extracted with traditional recommendation models as shown in Figure~\ref{fig:intro}(c). A set of base LoRAs can be fine-tuned using the grouped data and represented with corresponding user features. For each user, the engagement level of each LoRA can be determined through a gating mechanism. The difficulty of thinking is reduced through information classification and personalization mechanisms. 
In deployment, the reasoning ability internalized in model parameters allows ThinkRec to operate efficiently in the ranking stage: by constraining the output to a single token, the model produces batch-level results with just one forward pass.

We conduct experiments on three real-world recommendation datasets, validating the rationality and effectiveness of ThinkRec. ThinkRec average outperforms state-of-the-art baselines by 7.96\% in AUC and by 56.54\% in METEOR. In summary, the main contributions of this work are fourfold:
\begin{itemize}
\item We propose ThinkRec, a novel framework that shifts LLM-based recommendation from intuitive System 1 matching toward rational System 2-like reasoning, enabling more faithful and interpretable user preference modeling.
\item We design a data augmentation and reasoning activation framework that bridges recommendation with language modeling, leveraging metadata enrichment and reasoning guidance to trigger LLM reasoning in recommendations.
\item We introduce an instance-wise expert fusion mechanism that generates personalized recommendation experts based on latent user features, enhancing reasoning diversity and accuracy to handle diverse user behaviors.
\item Extensive experiments on three real-world datasets validate the effectiveness and reasonableness of ThinkRec.
\end{itemize}

\section{Related Work}
\subsection{LLM-based Recommendation}
\noindent \textbf{LLM-based Recommendation.}
With the rapid progress of LLMs, there has been growing interest in adapting them to recommendation tasks using both textual and structured data. Early works \cite{bao2023tallrec,ho2023btrec,lin2024rella,zhang2025collm} rely on prompt-based scoring by converting recommendations into binary question-answering, while others \cite {ji2023gen,li2023prompt,lin2024bridg,zhou2025onerectechnicalreport} align language space with item embeddings to directly generate item IDs. More advanced approaches, such as P5 \cite{geng2022p5} and InstructRec \cite{zhang2024agentcf} unify multiple sub-tasks—like rating prediction, pairwise comparison, or ranking, into a single language modeling framework. These methods demonstrate LLMs’ flexibility in expressing recommendation semantics, yet most focus on adapting output formats, treating LLMs as static scorers or selectors with limited interpretability. 
Beyond static prompting, recent efforts shift toward viewing LLMs as interactive agents capable of multi-step reasoning and personalized decision-making. For example, RecMind \cite{wang2024recmind} and MACRec \cite{wang2024macrec} introduce modularized reasoning structures, enabling planning-based recommendation. Others explore integrating LLMs into various stages of the pipeline—for instance, generating user or item representations \cite{zhang2024llmpower} or acting as high-level decision controllers \cite{park2025agentrec}. Meanwhile, instruction tuning \cite{zhang2024agentcf}, chain-of-thought prompting \cite{liu2025improving} have been proposed to enhance LLMs' semantic understanding. Despite these developments, explicit reasoning supervision and fine-grained user modeling remain underexplored. ThinkRec addresses these gaps by integrating reasoning-augmented training and personalized expert fusion, enabling user-specific recommendations grounded in structured reasoning.

\noindent \textbf{Reasoning Model.}
Recent advances have transformed LLMs from passive token predictors into structured reasoning agents by incorporating process-level supervision. Techniques such as Chain-of-Thought prompting \cite{2022weichain}, ReAct \cite{yao2022react}, and Tree-of-Thoughts \cite{yao2023tree} enable models to generate intermediate reasoning steps, improving interpretability and multistep inference quality. To further reinforce reasoning capabilities, recent works have introduced process reward models \cite{zhang2024rest} and self-improvement pipelines using techniques like Monte Carlo Tree Search \cite{luo2024improve} and reasoning distillation \cite{xu2023wizard}.
These trends define the emerging paradigm of Large Reasoning Models, which prioritize explicit, verifiable thinking processes over direct output. ThinkRec draws inspiration from this direction by introducing reasoning-augmented training for recommendation, aligning to embed structured reasoning into decision making.

\section{Method}
\begin{figure*}[t]
\centering
  \includegraphics[width=0.96\textwidth]{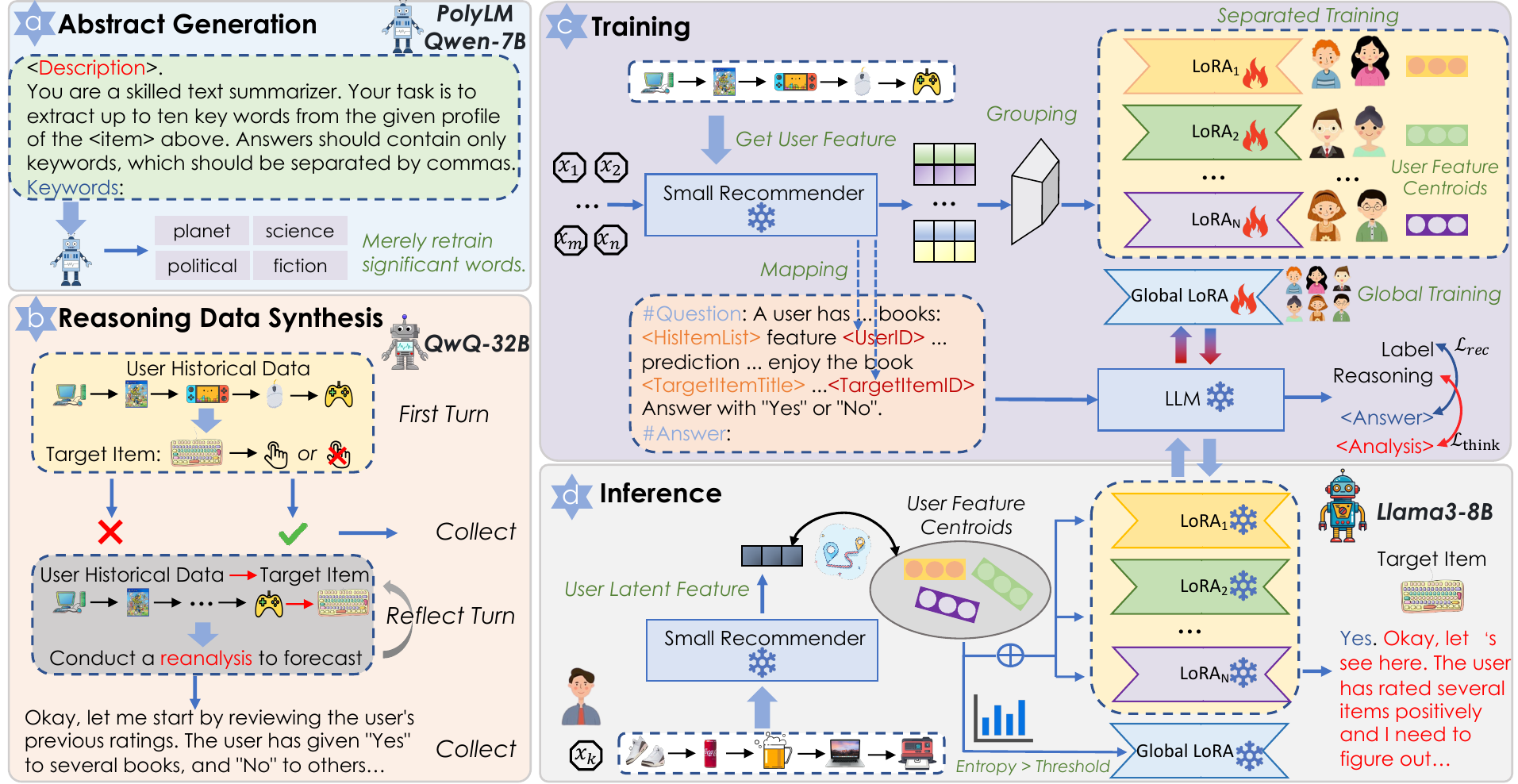}
\vspace{-0.3cm}
  \caption{Overview of the proposed ThinkRec. Keywords are extracted from the description of items with the pretrained PolyLM-Qwen-7B. Reasoning data is synthesized with a reasoning model from a sample of training data. Global LoRA is trained with total data, and base LoRAs are tuned on it with data grouping by user feature. The used LoRA, fusioned or single, is determined by the similarity between the user feature and LoRA representations.}
  \Description{Overview of the proposed ThinkRec. }
  \label{fig:method}
\vspace{-0.3cm}
\end{figure*}
\subsection{Preliminary}
\textbf{Problem Definition.} 
In the view of sequential recommendation, datasets can be formulated as $\mathcal{S}=\{(x_{u,t},y_{u,t})\}_{u=1,2,...,N,t=1,2,...,T_{u}}$, where $x_{u,t} = \{y_{u,1:(t-1)} \}$ denotes a user’s historical behaviors prior to the $t$th behavior $y_{u,t}$ and arranged in a chronological order, and $T_u$ denotes the number of behaviors for the user $u$. We will drop the sub-scripts occasionally and write $(x, y)$ in place of $(x_{u,t},y_{u,t})$ for simplicity. Each behavior $y$ is represented as $(u,t,i_{id},l)$, where $i_{id}$ is an item ID and $l \in \{1,0\}$ indicating the interaction label.
Furthermore,  additional textual information about items is provided, mainly the item title and description. Given $x_{u,t}$, get the textual information, and label to construct a historical text prompt to determine whether or not you would enjoy an item composed of the title and description of $y_{u,t}$ and give a reason $r_{u,t}$. We aim to enable the LLM to provide the underlying thinking beyond merely making recommendations as in previous work, thus achieving more effective and interpretable recommendations.
\\
\textbf{Integrating Collaborative Embeddings into LLMs.}
The sequential recommenders, such as MF \cite{koren2009mf} and LightGCN \cite{he2020lightgcn}, are used for encoding collaborative information, which can be integrated into LLM embeddings. Formally, for each sample $(u,t,i_{id},l)$:
\begin{equation}
    \mathbf{e}_{s}^{u}=f_{\psi}(u;\mathcal{S}); \mathbf{e}_{s}^{i}=f_{\psi}(i_{id};\mathcal{S}),
\end{equation}
where $\mathbf{e}_{s}^{u}\in \mathbb{R}^{1\times d1}$ denotes the user’s representation with dimension $d1$, $f_{\psi}(u;\mathcal{S})$ denotes the process used to obtain this representation, similarly for $i_{id}$. Formally, for an item $i$ with the text metadata $i_{txt}$, we obtain its textual tokens $\mathbf{E}_{txt}$ as follows:
\begin{equation}
    \mathbf{E}_{txt} = \mathrm{WE}(\mathrm{TKZ}(i_{txt})),
\end{equation}
where $\mathrm{TKZ}(\cdot),\mathrm{WE}(\cdot)$ present the LLM tokenizer and word embedding layer, respectively. $\mathbf{E}_{txt}=\{\mathbf{emb}_{txt}^{1:L}\}$, where $L$ is the number of tokens and $\mathbf{emb}_{txt}^{1}\in\mathbb{R}^{1\times d_2}$ denotes embeddings in the language space with dimension $d_2$. To align collaborative embeddings $\mathbf{e}_{s}^{i},\mathbf{e}_{s}^{u}$ to LLMs' language space, projector $proj_{\phi}(\cdot)$ is used:
\begin{equation}
    \mathbf{emb}_{s}^{i} = proj_{\phi}(\mathbf{e}_{s}^{i}),
    \mathbf{emb}_{s}^{u} = proj_{\phi}(\mathbf{e}_{s}^{u}).
\end{equation}

\subsection{Thinking Enhanced Recommendation}
To align recommendation with language modeling tasks, we highlight the importance of thinking activation for LLMs and propose a framework that consists of data construction and jointly training.
\subsubsection{Reasoning data construction}
To support LLMs for more effective reasoning, we need more textual information in addition to item titles. Given that the raw metadata often contains excessive or noisy content, directly using long-form text may hinder the understanding and inference of LLMs. To address this, we leverage a pre-trained summarization model \cite{wei2023polylm} to extract key concepts from the item metadata, and retain up to 10 representative keywords to enhance the semantic representation of each item. To model more comprehensive preference information, not just liking information, we add low-scoring interactions to the history sequence as well. Then we augment each item with title $i_{t}$, label $i_{l}\in\{yes,no\}$, feature, and keywords $i_{k}$ as follows:
\begin{tcolorbox}[colback=red!10!green!40!blue!10!white,colframe=red!20!green!40!blue]
$i_{txt}$ = $i_{t}$ with feature $\mathbf{emb}_{s}^{i}$ (label: $i_l$) with description: $i_{k}$.
\end{tcolorbox}
Since existing recommendation datasets lack explicit reasoning traces, we synthesize a subset of samples with explainable rationale by leveraging a powerful reasoning model QwQ\footnote{https://qwenlm.github.io/blog/qwq-32b/}. Specifically, we sample a few thousand instances from the training data and construct textual prompts according to Appendix~\ref{sec:prompt4sys}. As in Figure~\ref{fig:method}(b), we then repeatedly query the model until a correct prediction is produced and record the latest explanation content as the reason.
\subsubsection{Thinking activation}
\label{sec:think}
To equip the recommendation model with explicit reasoning capability, we introduce a thinking activation mechanism that jointly trains the model on both user–item interactions and synthesized reasoning examples. During training, we perform mixed sampling from conventional recommendation data with binary labels, and reasoning data generated by prompting a reasoning model to produce natural language explanations for user preferences. For each sample, we concatenate the input prompt and its corresponding answer ($i_{i}$ or $r_{u,t}$) to form a language modeling instance as input for training:
\begin{equation}
    \begin{array}{rl}
    \mathbf{E}^{qa}=&\mathrm{Concat}(\mathbf{E}^{q},\mathbf{E}^{a}),\\
    pos =& -\mathrm{Length}(\mathbf{E}^{a}),
    \end{array}
\end{equation}
where $\mathbf{E}^{q},\mathbf{E}^{a}$ means embedding of question and answer constructing from $\mathcal{S}$, respectively. While $pos$ marked the position of the first answer token. This mixed sampling strategy encourages the model to jointly learn the prediction of the recommendation outcome and the semantic alignment of the reasoning behavior, within a shared representation space, as shown in Figure~\ref{fig:method}(c).

The overall objective of training optimization consists of two loss components. For thinking instances, the standard token-level cross-entropy loss over the input sequence is used:
\begin{equation}
     \mathcal{L}_{think},logits=\mathrm{LLM}_{\theta}(\mathbf{E}^{qa}).
\end{equation}
For recommendation instances, we use the standard binary cross-entropy (BCE) loss between the predicted relevance score $\hat{l}$ and the observed ground-truth label $l$ of $y$:
\begin{equation}
\begin{array}{rl}
     posid=&\mathrm{TKZ}(\mathrm{'Yes'}),\\
     \hat{l}=&logits[pos][posid],\\
     \mathcal{L}_{rec}=&\mathrm{BCE}(\hat{l},l).
\end{array}
\end{equation}
To jointly optimize two objectives within a unified batch, we always use both losses simultaneously, adjusting the weight of each loss according to the instance type:
 \begin{equation}
     \mathcal{L} = \left\{\begin{array}{rl}
          \alpha\mathcal{L}_{rec}+\beta\mathcal{L}_{think},& thinking  \\
          \eta\mathcal{L}_{rec}+\gamma\mathcal{L}_{think},& recommend
     \end{array}\right..
 \end{equation}
 $\alpha,\beta,\eta,\gamma$ are weights for each loss. The model learns not only to rank relevant items but also to generate grounded reasons for its predictions, effectively activating the model's thinking ability.
\subsection{Recommendation Experts Fusion}
\subsubsection{Base Expert fine-tuning}
To enable each expert to better capture user-specific preferences and behavioral patterns, we partition users into distinct groups based on latent features. Specifically, we utilize user embeddings derived from various pretrained small collaborative models\cite{lv2025collaboration,fu2024diet,zhang2022latentstructureminingcontrastive}, each of which implicitly encodes the interaction semantics of users. These embeddings serve as the basis for grouping. We aggregate the representations across all users and perform unsupervised clustering to obtain $N$ groups $\mathcal{S}^{\prime}=\{\mathcal{S}_{1:N}\}$. The resulting clusters are then used to partition the training and validating data, allowing each expert to specialize in a subset of users with similar representations to simplify the preference modeling.

We first fine-tune a global expert using all the data in the framework described in Section~\ref{sec:think}, employing LoRA-based adaptation to enable generalizable recommendations with thinking\cite{lv2023duet,lv2025tackling,ijcai2025p402}. Building on this global expert $\mathrm{LoRA}_{global}$, we further fine-tune the selected LoRA layers (the last 8 layers) with grouped data, enabling the model to preserve general thinking capability while adapting to finer-grained user preferences in recommendation.
This produces a set of base experts $\{\mathrm{LoRA}_{1:N}\}$ that act as foundations for user-specific dynamic expert fusion.

\subsubsection{Instance Wise Expert Fusion}
To determine which expert is most suitable for a given user, we assign the mean of user features extracted within each group by the corresponding small model as representations of experts $\mathcal{E}=\{\mathbf{e}_{1:N}^{c}\}$. We then compute the match between user features and expert representations to estimate each base expert’s involvement in modeling the user’s preferences. The cosine similarity and softmax functions were used to obtain participation scores $\mathbf{w}^{u}=\{w^{u}_{1:N}\}$ of user $u$ based on experts:
\begin{equation}
\begin{array}{rl}
    \mathbf{e}_{n}^{c} =& \mathrm{Mean}(\mathbf{e}^{u}_{s}),u \in \mathcal{S}_{n},\\
    \mathbf{w}^{u}=&\mathrm{Softmax}(\mathrm{Cosim(\mathbf{e}_{s}^{u},\mathcal{E})}/\tau),
\end{array}
\end{equation}
where $\tau$ is the temperature coefficient. We introduce a gating mechanism to filter users with highly averaged $\mathrm{H}(\mathbf{w}^{u}) > 0.95\log N$ or concentrated $\max(\mathbf{w}^{u})>0.5+0.6/N$ preference profiles, assigning them directly to a global or base expert; the remaining users are served by instance-wise fusion over multiple experts as shown in Figure~\ref{fig:method}(d). The threshold is calculated as follows:
\begin{equation}
\mathrm{H}(\mathbf{w}^{u}) = -\sum_{n=1}^{N}w_{n}^{u}\log{w_{n}^{u}},
\end{equation}
\begin{equation}
\mathrm{LoRA}^{u} = \left\{\begin{array}{ll}
     \mathrm{LoRA}_{global},& \mathrm{averaged} \\
     \mathrm{LoRA}_{argmax(\mathbf{w}^{u})},& \mathrm{concentrated} \\
     \sum_{n=1}^{N}w_{n}^{u}\mathrm{LoRA}_{n},& \mathrm{otherwise}
\end{array} \right..
\end{equation}

\subsection{Staged Training}

In this section, we describe the training process. First, we train a global LoRA $\mathrm{LoRA}_{global}$ on the full dataset under pure text conditions. In parallel, we train a collaborative model $f_{\psi}(\cdot;\mathcal{S})$ on the full dataset under ID-based conditions. Next, user features extracted from the collaborative model are used to group the data, and the grouped data are employed to fine-tune the last 8 layers of $\mathrm{LoRA}_{global}$, yielding a set of base LoRAs $\{\mathrm{LoRA}_{1:N}\}$. Finally, leveraging the global dataset with injected collaborative signals, we train a projector $proj_{\phi}(\cdot)$ that projects collaborative embeddings into the LLM language space, while keeping $\mathrm{LoRA}_{global}$ fixed.

\section{Experiments}
We conduct experiments on real-world datasets to answer three main research questions: \textbf{RQ1:} How does ThinkRec perform in comparison to existing recommendation methods? \textbf{RQ2:} Why is the thinking activation method in ThinkRec essential? \textbf{RQ3:} How does the fusion of experts influence recommendation performance?

\subsection{Experimental Setup}
\subsubsection{Datasets.}
We conduct experiments on three datasets: \textit{ML1M} (MovieLens-1M)\footnote{https://grouplens.org/datasets/movielens/1m/}, \textit{Yelp} (Yelp Open Dataset)\footnote{https://business.yelp.com/data/resources/open-dataset/}, and \textit{Book} (the “Book” subset of the Amazon Product Review dataset)\footnote{https://nijianmo.github.io/amazon/index.html}. To better simulate real-world recommendation scenarios and avoid data leakage \cite{Ji2023critical}, we split each dataset into training, validation, and testing sets according to interaction timestamps. Specifically, for Amazon-Book, due to its large scale, we retain interactions from 2017 only, using the first 10 months for training and the last 2 months for validation and testing. For Yelp, we retain interactions from 2010 to 2022, assigning the first 10 years to training and the last 2 years to validation and testing. For ML1M, we keep interactions from the most recent 20 months, with the first 10 months for training and the remaining 10 months for validation and testing. To address sparsity in Book and Yelp, we filter out users and items with fewer than 20 interactions. The statistics of the processed datasets are summarized in Table~\ref{tab:static}, where $\mu$ and $\sigma$ denote the mean and standard deviation of user interactions, respectively.

Regarding labels, \textit{ML1M} contains user ratings on movies, collected between 2000 and 2003, with ratings on a scale of 1 to 5. We convert these ratings into binary labels using a threshold of 3. \textit{Yelp} includes user reviews, ratings for businesses such as restaurants and retail shops, as well as textual information about the businesses. We convert these ratings into binary labels using a threshold of 3. \textit{Book} compiles user reviews of books from Amazon, collected between 1996 and 2018, with review scores ranging from 1 to 5. We transform these review scores into binary labels using a threshold of 4. Ratings greater than the threshold are labeled as “positive” (y = 1), while the rest are labeled as “negative” (y = 0).
\begin{table}[tb]
  \caption{Statistics of Processed Datasets.}
  \label{tab:static}
  \vspace{-0.2cm}
  \centering
  \resizebox{\linewidth}{!}{ 
  \begin{tabular}{lcccccc}
    \toprule[2pt]
    \textbf{Datasets} & \textbf{\#Train} & \textbf{\#Valid} & \textbf{\#Test} & \textbf{\#User} & \textbf{\#Item} & $\mathbf{\sigma/\mu}$ \\
    \midrule
    \textbf{ML1M} & 33,891 & 10,401 & 7,331 & 5,945 & 3,687 &  1.48\\
    \textbf{Yelp} & 1,637,168 & 144,929 & 144,929 & 40,617 & 60,014 & 1.24\\
    \textbf{Book} & 650,865 & 56,262 & 56,262 & 22,686 & 47,059 & 1.07\\
    \bottomrule[2pt]
  \end{tabular}
  }
\vspace{-1em}
\end{table}

\begin{table*}[ht]
  \caption{Comparison of prediction performance between ThinkRec and the baselines across the three evaluation datasets. The best results are highlighted in \textbf{bold} and sub-optimal results are \underline{underlined}.}
  \label{tab:main}
  \vspace{-0.2em}
\setlength{\tabcolsep}{7pt}
\renewcommand{\arraystretch}{0.95}
\resizebox{1.0\textwidth}{!}{
\begin{tabular}{@{}c|cccc|cccc|cccc@{}}
\toprule[2pt]
\textbf{Datasets}  & \multicolumn{4}{c|}{\textbf{ML1M}}                                              & \multicolumn{4}{c|}{\textbf{Yelp}}                                              & \multicolumn{4}{c}{\textbf{Book}}                                              \\ 
\midrule
Methods    & AUC             & UAUC            & NDCG@5          & MAP@5           & AUC             & UAUC            & NDCG@5          & MAP@5           & AUC             & UAUC            & NDCG@5          & MAP@5           \\ \hline
MF        & 0.6401          & 0.6079          & 0.7286          & 0.4520          & 0.5838          & 0.5389          & 0.8120          & 0.2552          & 0.6592          & 0.5527          & 0.6805          & 0.2887          \\
LightGCN  & 0.6140          & 0.6230          & 0.7333          & 0.4600          & 0.5360          & 0.5179          & 0.8076          & 0.2520          & 0.5622          & 0.4985          & 0.6406          & 0.2598          \\
SASRec    & 0.6956          & 0.6687    & 0.7663          & \underline{0.4747}    & 0.6184          & \textbf{0.6096} & 0.8564    & 0.2785    & 0.5411          & 0.5197          & 0.6550          & 0.2701          \\ 
gSASRec    & 0.7043&\underline{0.6690}&0.7650&0.4693&0.6200&0.6051&\textbf{0.8586}&\underline{0.2792}&0.6187&0.5546&0.6858&0.2964\\ 
\midrule
Prompt4NR & 0.6936          & 0.6433          & 0.7665          & 0.4652          & 0.6272          & 0.6034          & 0.8348          & 0.2705          & 0.6764          & \underline{0.5699}    & \textbf{0.7023} & \underline{0.3048}    \\
TALLRec   & 0.6872          & 0.6553          & \underline{0.7683}    & 0.4706          & 0.5334          & 0.5206          & 0.7988          & 0.2538          & 0.6632          & 0.5568          & \underline{0.7023}    & \textbf{0.3049} \\
CoLLM     & \underline{0.7141}    & 0.6672          & 0.7585          & 0.4647          & \underline{0.6373}    & 0.5961          & 0.8420          & 0.2734          & \underline{0.7830}    & 0.5672          & 0.6917          & 0.2968          \\ \hline
\rowcolor[HTML]{BDDBED} 
Ours      & \textbf{0.7764} & \textbf{0.6775} & \textbf{0.7747} & \textbf{0.4774} & \textbf{0.6955} & \underline{0.6065}    & \underline{0.8585} & \textbf{0.2826} & \textbf{0.8302} & \textbf{0.5705} & 0.6858          & 0.2977          \\ 
\bottomrule[2pt]
\end{tabular}
}
\vspace{-0.3em}
\end{table*}
\subsubsection{Baselines.}
To fully evaluate the proposed method ThinkRec, we compare it with traditional methods and LLM-based methods:
\begin{itemize}
    \item \textbf{MF.} \cite{koren2009mf} This refers to Matrix Factorization, a representative latent factor-based collaborative filtering method. 
    \item \textbf{LightGCN.} \cite{he2020lightgcn} A representative graph-based collaborative filtering method, which uses a graph convolutional neural network to enhance the modeling of user interest. 
    \item \textbf{SASRec.} \cite{kang2018sasrec} A representative sequential recommendation method, which uses self-attention to encode sequential patterns to model user interest.
    \item \textbf{gSASRec.} \cite{gSASRec} A graph-enhanced sequential recommendation model that augments SASRec by incorporating item–item transition graphs to capture sequential dependencies.
    \item \textbf{Prompt4NR.} \cite{zhang2023prompt4nr} It uses both fixed and soft prompts to utilize traditional Language Models for recommendation. 
    \item \textbf{TALLRec.} \cite{bao2023tallrec} This is a state-of-the-art LLMRec method that aligns LLM with recommendations through aligns LLM with recommendations through instruction tuning.
    \item \textbf{CoLLM.} \cite{zhang2025collm} It effectively integrates collaborative information into LLMs by harnessing the capability of external traditional models to capture the information.
\end{itemize}
We extend the above LLM-based method to LLM Llama3-8B\footnote{https://huggingface.co/meta-llama/Meta-Llama-3-8B-Instruct} for a fair comparison and tune the LLM with LoRA to manage costs. 

\subsubsection{Evaluation Metrics.}
We employ four widely used recommendation metrics: \textit{AUC}, \textit{UAUC}, \textit{Normalized Discounted Cumulative Gain (NDCG)}, and \textit{Mean Average Precision (MAP)}.
And we employ \textit{METEOR} \cite{lavie2005meteor} and \textit{BLEURT} \cite{dan2020bleurt} to measure the generated reasons. METEOR incorporates synonym matching and word order. BLEURT is fine-tuned on human judgments to directly predict text quality.

\subsubsection{Implementation Details}
We implement all the compared methods using PyTorch 2.5. We adopt BCE when not otherwise specified. LLM-based methods are optimized using AdamW, while other models use Adam.
For hyperparameters such as learning rate, embedding size, and weight decay of traditional methods, we conduct grid search over commonly used ranges, and follow the original papers for other baseline-specific settings. For LLM-based methods, we adopt CoLLM's configuration for optimizer, sequence truncation and LoRA settings. Full hyperparameter settings and search ranges are provided in the Appendix~\ref{sec:Details}.

\subsection{Overall Performance (RQ1)}
\subsubsection{Accuracy of Recommendation}
The results of ThinkRec and SOTA recommendations are in Table~\ref{tab:main}. We observe improvement in our method on all three datasets. Notably, ThinkRec improves the previous SOTA CoLLM by +.0582 (9.13\%) on AUC in Yelp and +.0623 (8.72\%) on AUC in ML1M, demonstrating substantial gains in global ranking accuracy. CoLLM, which integrates collaborative signals into LLMs using external traditional models, achieves the second-best AUC on ML1M (0.7141) and Book (0.7830), confirming its effectiveness in leveraging user–item interaction patterns. However, its performance in user-level ranking metrics is relatively less competitive, especially on ML1M and Yelp, where ThinkRec offers more personalized modeling through instance-wise expert fusion.

Other LLM-based methods, such as Prompt4NR and TALLRec, also demonstrate competitive performance, particularly on datasets with rich textual item metadata such as Book. For instance, TALLRec achieves the highest NDCG@5 (0.7683) on ML1M among all methods except ThinkRec, and both TALLRec and Prompt4NR slightly outperform ThinkRec in MAP@5 on Book. These results suggest that instruction tuning and prompt engineering are beneficial when item descriptions provide substantial semantic context.

Turning to traditional recommendation models, SASRec clearly outperforms MF and LightGCN, especially on ML1M and Yelp. Its self-attention-based sequential modeling effectively captures temporal patterns in user behavior, yielding the best UAUC (0.6687) and MAP@5 (0.4747) among non-LLM baselines on ML1M. However, its performance declines significantly on the Book dataset, where user interactions are sparser and exhibit weaker sequential patterns, revealing the inherent limitations of models that rely solely on sequential behavior modeling.

Overall, ThinkRec delivers the most balanced and robust performance across all datasets and evaluation metrics. It achieves the best performance of almost all metrics on ML1M and Yelp, and also obtains the top AUC and UAUC on Book. Its consistent top-tier results confirm the effectiveness of combining thinking activation and instance-wise expert fusion. These components jointly enhance both global and user-specific ranking quality, making ThinkRec a scalable and interpretable recommendation.

\subsubsection{Quality of Generated Reasons}
Table~\ref{tab:quality} summarizes the quality of the reasons generated by the LLM-based recommenders. Compared with the generated reasons from QwQ using the METEOR and BLEURT metrics, the reasons generated by our method significantly outperform those of the three LLM-based baselines. Our method achieves an average relative improvement of 56.54\% on METEOR and 23.35\% on BLEURT across all datasets, suggesting better fluency, coherence, and semantic relevance. These results validate the effectiveness of our thinking activation mechanism, which explicitly aligns recommendations with user-centric reasoning via joint training on reasoning-augmented samples. The improvement in both syntactic and semantic metrics confirms that ThinkRec not only provides accurate recommendations but also produces more coherent, grounded, and human-aligned explanations, a crucial step toward reasonable and trustworthy LLM-based recommendations.
\begin{table}[tb]
\caption{Quality evaluation of generated reasons. "M" refers to "METEOR" and "B" refers to "BLEURT".}
\label{tab:quality}
\renewcommand{\arraystretch}{0.95}
\resizebox{\linewidth}{!}{
\begin{tabular}{c|cc|cc|cc}
\toprule[2pt]
\textbf{Datasets}  & \multicolumn{2}{c|}{\textbf{ML1M}}          & \multicolumn{2}{c|}{\textbf{Yelp}}          & \multicolumn{2}{c}{\textbf{Book}}          \\ 
\midrule
Methods    & M          & B          & M          & B          & M          & B          \\ 
\midrule
Prompt4NR & 0.0010          & 0.2013          & 0.0205          & 0.1675          & 0.0003          & \underline{0.1957}    \\
TALLRec   & \underline{0.0275}    & \underline{0.2607}    & \underline{0.0379}    & \underline{0.2420}    & \underline{0.0301}    & 0.1931          \\
CoLLM     & 0.0003          & 0.1626          & 0.0001          & 0.1785          & 0.0097          & 0.1636          \\ 
\midrule
\rowcolor[HTML]{BDDBED} 
Ours      & \textbf{0.0333} & \textbf{0.3104} & \textbf{0.0616} & \textbf{0.2683} & \textbf{0.0546} & \textbf{0.2828} \\ 
\bottomrule[2pt]
\end{tabular}}
\vspace{-0.8em}
\end{table}

\begin{figure}[tb]
  \includegraphics[width=\columnwidth]{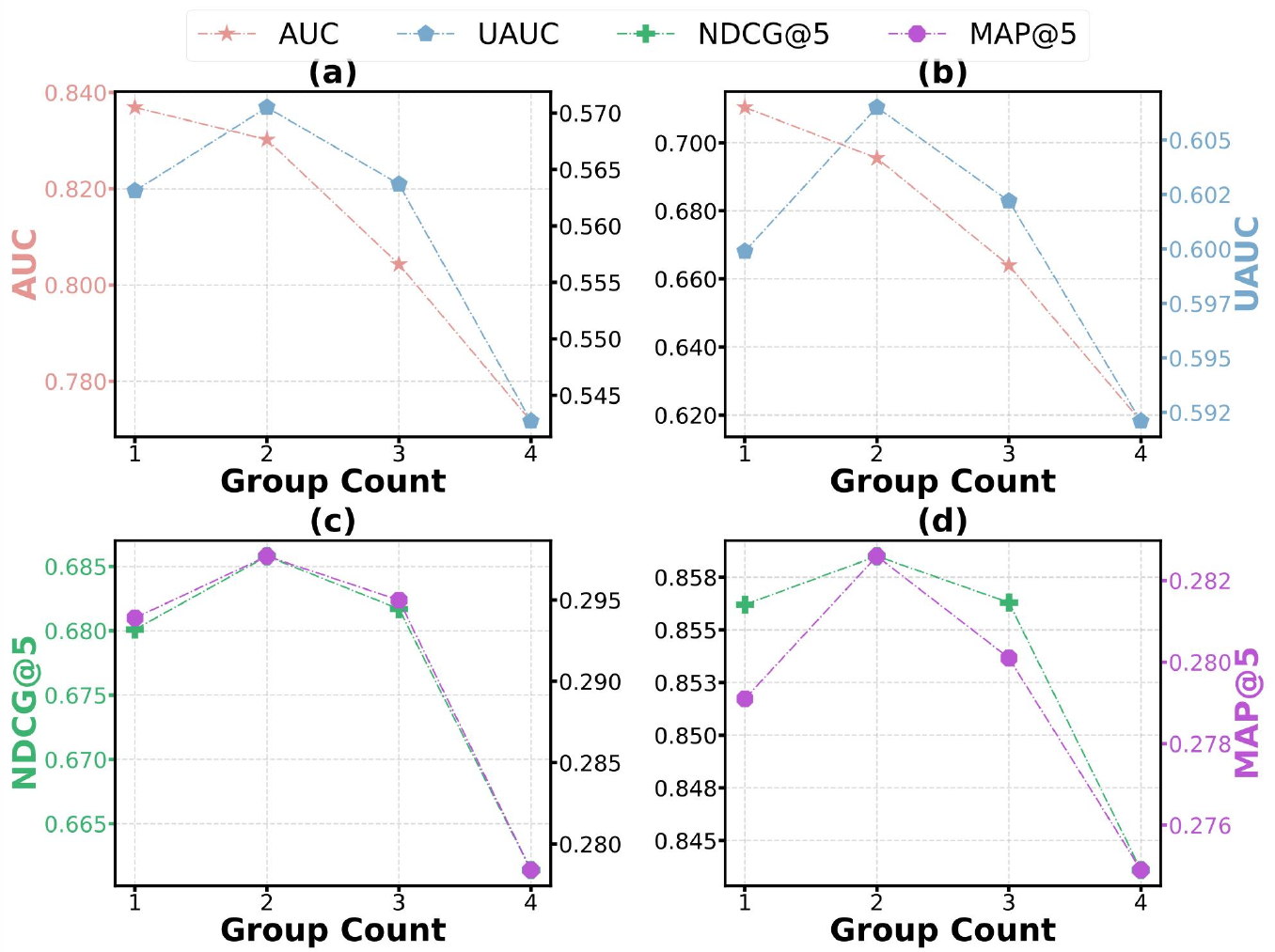}
  \caption{The influence of performance with the number of experts on Book (left panel) and Yelp (right panel).}
  \Description{The influence of performance with the number of experts on Book (left panel) and Yelp (right panel).}
  \label{fig:groups}
\vspace{-0.6em}
\end{figure}

\subsubsection{Case Study}
Among existing LLM-based recommendations, CoLLM and Prompt4NR yield disorganized symbols. TALLRec frequently generates sentences with unrelated elements, such as code or hallucinated facts, failing to reflect coherent reasoning. In contrast, ThinkRec demonstrates structured, step-by-step reasoning aligned with user history and target item semantics, enabling it to produce accurate and interpretable recommendations. A real example is listed in the Appendix~\ref{sec:case}.

\subsection{In-depth Analysis}
\subsubsection{Ablation Studies (RQ2)}
To evaluate the importance of explicit reasoning in recommendation, we ablate the "thinking" component of ThinkRec, which disables reasoning supervision (w/o think). As shown in Table~\ref{tab:ablation}, this leads to significant performance degradation in all datasets. For example, UAUC on the Book dataset drops from 0.5705 to 0.4692. Interestingly, this is even lower than the variant in which both thinking and expert mechanisms are removed (w/o both). 
This counterintuitive result suggests that, in the absence of thinking, the recommendation task degenerates into binary classification, where multi-expert modeling risks overfitting superficial interactions and thereby undermining generalization.
 
We then assess the contribution of the expert personalization module, which removes the latent-feature-based user grouping and experts fusion mechanism (w/o experts). As shown in Table~\ref{tab:ablation}, this also leads to consistent performance drops—for instance, MAP@5 on Yelp falls from 0.2826 to 0.2791. Notably, only when thinking is enabled does multi-expert modeling begin to show substantial benefits. With reasoning supervision, group-specific LoRA modules can effectively specialize in distinct user groups, capturing fine-grained preference signals that would otherwise be blurred in the global model. These findings highlight the consistency and complementarity between thinking and multi-expert modeling, providing a semantically rich space that allows user grouping to generalize rather than overfit, enabling expert models to move beyond surface interaction patterns and capture deeper preference semantics.
\begin{table}[tb]
\caption{Ablation studies of key components in ThinkRec. "N" refers to "NDCG", "M" refers to "MAP".}
\label{tab:ablation}
\renewcommand{\arraystretch}{0.95}
\setlength{\tabcolsep}{1pt}
\resizebox{\linewidth}{!}{
\begin{tabular}{c|ccc|ccc|ccc}
\toprule[2pt]
\textbf{Datasets}     & \multicolumn{3}{c|}{\textbf{ML1M}}                            & \multicolumn{3}{c|}{\textbf{Yelp}}                            & \multicolumn{3}{c}{\textbf{Book}}                            \\ 
\midrule
Methods     & UAUC            & N@5          & M@5           & UAUC            & N@5          & M@5           & UAUC            & N@5          & M@5           \\ 
\midrule
w/o both    & 0.6658          & 0.7643          & 0.4674          & 0.5904          & 0.8429          & 0.2736          & 0.5017          & 0.6381          & 0.2613          \\
w/o think   & 0.6599          & 0.7570          & 0.4623          & 0.5865          & 0.8402          & 0.2702          & 0.4692          & 0.6284          & 0.2548          \\
w/o experts & 0.6765    & 0.7740    & 0.4742    & 0.5999    & 0.8562    & 0.2791    & 0.5631    & 0.6801    & 0.2939    \\
\midrule
\rowcolor[HTML]{BDDBED}
Ours        & \textbf{0.6775} & \textbf{0.7747} & \textbf{0.4774} & \textbf{0.6065} & \textbf{0.8585} & \textbf{0.2826} & \textbf{0.5705} & \textbf{0.6858} & \textbf{0.2977}
\\ 
\bottomrule[2pt]
\end{tabular}}
\vspace{-0.3em}
\end{table}
\begin{table}[tb]
\caption{Metrics before and after model parameter updates.}
\label{tab:multi}
\vspace{-0.2em}
\setlength{\tabcolsep}{11pt}
\resizebox{\linewidth}{!}{
\begin{tabular}{cccc}
\toprule[2pt]
\textbf{Models}              & UAUC            & NDGC@5          & MAP@5           \\
\midrule
global              & 0.5098          & 0.6462          & 0.2657          \\
global* & 0.5119 (0.41\%) & 0.6542 (1.24\%) & 0.2704 (1.77\%) \\
\midrule
2 experts           & 0.5380          & 0.6641          & 0.2805          \\
3 experts           & 0.5669 (5.37\%) & 0.6890 (3.75\%) & 0.2950 (5.17\%)\\
\bottomrule[2pt]
\end{tabular}
}
\vspace{-0.8em}
\end{table}
\subsubsection{Study on the Fusion of Experts (RQ3)}
\paragraph{Necessity of multiple experts.}
As shown in Figure~\ref{fig:groups}, ThinkRec achieves clear improvements at the optimal number of expert groups (2–3), particularly in UAUC. Prior works\cite{2017DeepFM,2018DIN} emphasize UAUC as a key performance indicator, where even small gains are considered valuable in practice. In real-world scenarios.
Moreover, the multi-expert design improves robustness by capturing diverse user behaviors and enables modular expansion without retraining the full model.  We divided the Book dataset into three groups: two groups of original data and one group of newly introduced data. We then tested the performance of the unified and multi-expert models separately. As shown in Table~\ref{tab:multi}, updating only the unified model parameters provides minimal improvement as the number of users increases. In contrast, the multi-expert model has higher original metrics and shows greater improvement.

\paragraph{Analysis of the number of experts.}
As shown in Figure~\ref{fig:groups}, the number of expert groups increases from 1 to 4, the model exhibits a characteristic 'rise-then-fall' performance trend, revealing the trade-off between personalization capacity and generalization. In the early stages, fine-tuning LoRA modules within user groups significantly enhances the model’s ability to capture diverse preferences, resulting in notable gains in user-level and Top-N ranking metrics such as UAUC, NDCG@5, and MAP. However, with further partitioning, each subgroup receives fewer training samples, making the model prone to overfitting, thereby degrading ranking performance. Notably, the AUC metric consistently decreases with more experts, reflecting the deterioration of the consistency of global representation with expert specialization, validating the inherent tension between "global consistency" and "local specificity" in recommender systems. These results indicate that more experts do not necessarily equate to better performance; instead, the optimal group number should be dynamically adjusted according to user behavior diversity, and the frequency of interactions.
\begin{figure}[tb]
\centering
  \includegraphics[width=0.9\columnwidth]{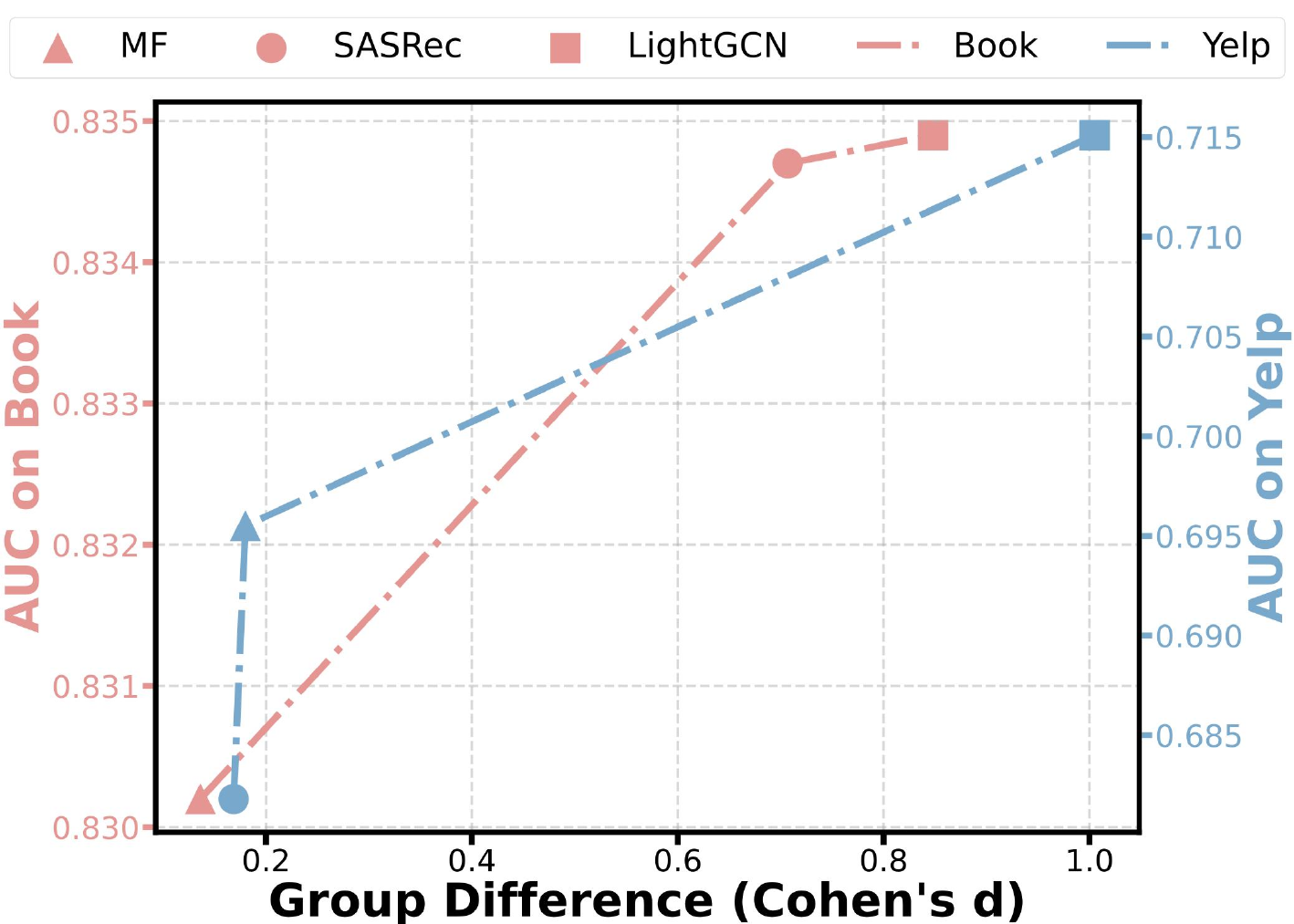}
  \caption{The influence of performance on the accuracy of grouping (Cohen's d of grouped datasets).}
  \Description{The influence of performance on the accuracy of grouping (Cohen's d of grouped datasets).}
  \label{fig:groups_m}
\vspace{-0.7em}
\end{figure}
\\
\paragraph{Analysis of the grouping features.}
Under a fixed two-group setting, we further investigate how the choice of user grouping features affects model performance. We employ user embeddings generated by MF, LightGCN, and SASRec to construct different grouping strategies. As shown in Figure~\ref{fig:groups_m}, with the increase of group difference (Cohen’s d), the performance of ThinkRec consistently improves. This trend highlights the importance of semantic decoupling among expert groups. When user preferences across groups exhibit stronger heterogeneity, the LoRA modules assigned to each group can learn more complementary preference representations, thereby enhancing the system’s modeling capacity and global discriminative power. Therefore, leveraging high-quality user behavior modeling methods as the basis for grouping can amplify divergence across groups, enabling multi-expert systems to achieve better personalized expressiveness while preserving a global perspective.
\subsubsection{Cross‑Domain Generalization}
In practical recommender systems, distinct models are often deployed for different domains due to significant variations in user behavior across scenarios. In multi-domain settings, a single model is typically trained on a mixture of item types to handle all cases. In contrast, ThinkRec introduces a more nuanced strategy: each base expert is trained on data grouped by user latent features that are specific to the target domain, allowing the model to capture fine-grained, domain-specific preference patterns while maintaining a unified reasoning backbone. To rigorously evaluate cross-domain generalization, we conducted experiments using the global expert. Specifically, we test the metrics in the datasets themselves and use experts trained on the ML1M dataset to test the metrics on the Book and Yelp datasets, as shown in Table~\ref{tab:general}. The results demonstrate that the reasoning paths learned from ML1M transfer effectively to other domains, retaining more than $90\%$ of the performance achieved with domain-specific training. This demonstrates ThinkRec's strong generalization capacity and its ability to leverage shared reasoning structures across diverse web environments, reducing the need for domain-specific retraining while maintaining robust recommendation quality.
\begin{table}[tb]
\caption{Cross‑Domain Generalization Evaluation.}
\label{tab:general}
\resizebox{\linewidth}{!}{
\begin{tabular}{cccccc}
\toprule[2pt]
\textbf{Datasets} & ML1M   & Yelp   & ML1M->Yelp & Book   & ML1M->Book \\
\midrule
AUC      & 0.7843 & 0.7104 & 0.6709 (94.4\%)         & 0.8369 & 0.8001 (95.6\%)         \\
NDGC@5   & 0.7740 & 0.8562 & 0.7717 (90.1\%)         & 0.6801 & 0.6197 (91.1\%)         \\
METEOR   & 0.0970 & 0.0892 & 0.0876 (98.2\%)         & 0.1028 & 0.1040 (101.2\%)        \\
BLEURT   & 0.4043 & 0.3491 & 0.3421 (98.0\%)         & 0.3489 & 0.3562 (102.1\%)      \\
\bottomrule[2pt]
\end{tabular}
}
\end{table}

\subsubsection{Inference Efficiency}
Although ThinkRec leverages multiple LoRA‑tuned experts during training, inference requires only a single LLM forward pass with the fused LoRA adapter (via weighted sum of a few small matrices),
and a lightweight gating computation over user embeddings.
Because LoRA adapters add only a small fraction ($< 1.5\%$) of extra parameters and incur negligible runtime overhead, the overall latency increase is minimal (empirically $< 10\%$ relative to a standard LoRA‑tuned LLM) while delivering substantially richer, interpretable outputs. 

\section{Conclusion}
The core contribution of this research is advancing LLM-based recommendation systems from an intuitive, matching-based "System 1" paradigm to a "System 2" paradigm characterized by explicit reasoning. Through our proposed thinking activation mechanism and instance-wise expert fusion strategy, ThinkRec deeply couples the recommendation task with the reasoning capabilities of language models. Experimental results confirm that this approach not only significantly improves recommendation accuracy but, crucially, enhances system transparency and user trust by generating traceable reasons.
We believe that future research will move towards deeper cognitive insights or exploring training-free mechanisms like activation steering for personalized reasoning\cite{zhang-etal-2025-personalized}, enabling systems to perform more rigorous logical reasoning. Moreover, based on the principles of automatic and personalized selection, deeper levels of Collective Intelligence, such as in Multi-Agent Systems, can be explored. ThinkRec builds a solid bridge for this leap. We provide the ethical use of data in Appendix~\ref{sec:eth}.
\begin{acks}
This work was supported by National Natural Science Foundation of China (No. 62402429, U24A20326, 62441236), the Key Research and Development Program of Zhejiang Province (No. 2025C01026, 2024C03270), Ningbo Yongjiang Talent Introduction Programme (2023A-397-G), Young Elite Scientists Sponsorship Program by CAST (2024QNRC001), and partially supported by MYbank, Ant Group.
The author gratefully acknowledges the support of Zhejiang University Education Foundation Qizhen Scholar Foundation.
\end{acks}
\newpage

\bibliographystyle{www26}
\bibliography{www26}

\appendix
\section{Ethical Considerations}
\label{sec:eth}
We use only publicly available benchmark datasets (e.g., ML1M, Yelp, Book) that are widely adopted in the recommendation community. These datasets do not contain any personally identifiable information (PII) or sensitive attributes, and our study does not involve human participants or subjects, thus not requiring informed consent. All data usage strictly complies with ACM’s Publications Policy on Research Involving Human Participants and Subjects.
\section{Prompt Templates}
\subsection{Prompt for Summarizing Metadata}
\begin{tcolorbox}[colback=red!10!green!40!blue!10!white,colframe=red!20!green!40!blue]
\textbf{Summarization}\\
Your task is to extract up to ten keywords from the given profile of the book above. Answers should contain only keywords, which should be separated by commas.\\Keywords:
\end{tcolorbox}
\subsection{Prompt for Reasoning Data Synthesis}
\label{sec:prompt4sys}
\begin{tcolorbox}[colback=red!7!white,colframe=red!40!white]
\textbf{First Turn}\\
A user has given high ratings to the following books: \textit{<HisItemList>}. Using all available information, make a prediction about whether the user would enjoy the book titled \textit{<TargetItemTitle>}?
\end{tcolorbox}

\begin{tcolorbox}[colback=red!10!green!40!blue!10!white,colframe=red!20!green!40!blue]
\textbf{Reflect Turn}\\
$\rhd$ The correct response is \textit{<answer>}. Reflect on multiple aspects based on historical information and explain the reason for the oversight based on the previous analysis. Reanalyze to make a prediction about whether the user would enjoy the book titled \textit{<TargetItemTitle>}?\\
$\rhd$ The accurate answer is \textit{<answer>}. Delve into various aspects considering historical data, elucidate the cause of the oversight according to the preceding analysis. Conduct a reanalysis to forecast whether the user will take pleasure in the book named \textit{<TargetItemTitle>}?\\
$\rhd$ The right response is \textit{<answer>}. Reflect on a variety of aspects with reference to historical information, and account for the oversight based on the earlier analysis. Reanalyze to determine whether the user would appreciate the book titled \textit{<TargetItemTitle>}?
\end{tcolorbox}
\subsection{Prompt for Recommendation}
\label{sec:prompt4rec}
\begin{tcolorbox}[colback=red!7!white,colframe=red!40!white]
\textbf{Recommendation}\\
\#Question: A user has given ratings to the following books: \textit{<HisItemList>}. Additionally, we have information about the user's preferences encoded in the feature \textit{<UserID>}. Based on the descriptions and the user's enjoyment of each book in the historical sequence, construct a persona of the user's preferences and reevaluate whether the user would enjoy the book titled \textit{<TargetItemTitle>} with the feature \textit{<TargetItemID>}. Please begin your analysis with "Yes" or "No". \\
\#Answer:
\end{tcolorbox}

\section{Case Study}
\label{sec:case}
\begin{tcolorbox}[colback=red!10!green!40!blue!10!white,colframe=red!20!green!40!blue]
\textbf{Prompt4NR}\\
")) '"'); 3.\}"', '); \}') { }, \} "); \}; """ ``` I think I understand the problem, but it's a bit unclear to me. ") \}; \} """ ");
\end{tcolorbox}
\begin{tcolorbox}[colback=red!25!green!30!blue!5!white,colframe=red!20!green!40!blue]
\textbf{TALLRec}\\
");" \# 3. The user's response: "1"; (the number of the book to be analyzed) \# - If a book has been rated, then it is included in this list. \# * For each book that matches with any other book on the same list: \# + Check if both books are novels and have similar genres \# - In case they match these criteria, add them together as one single entry """ def predict\_book\_rating(book\_id): for i in range(len(user\_ratings)): if int(i) == book\_id: return True result = [] for j in user\_ratings: if len(result) > 0: flag = False for k in result[-1]: if j[book\_name] == k[0][book\_title]: if 'novel' in j['genres'] or 'fiction' or '': if ('romance', '') not in k: result.pop() break else: flag=True continue elif j.get('genres') != None and set(j['genre']) \& set(k[2]): flag=False \# \#!/usr/bin/env python from flask import Flask, request, jsonify import json app = Flask(\_\_name\_\_) @app.route('/predict-book-rating') def predict(): data = \{'user': ['The Duchess Deal: Girl Meets Duke'; 'The V Card';...]\} print(data) if \_\_name\_\_ == '\_\_main\_\_': app.run(debug=True) ``` Here you can use Python script. Please note that I am using `flask` framework which allows us to create simple web applications. Here we will make an API endpoint `/predict-books-ratings`. You should send JSON object where key `'books'` contains array of book titles.
\end{tcolorbox}

\begin{tcolorbox}[colback=red!10!green!40!blue!10!white,colframe=red!20!green!40!blue]
\textbf{CoLLM}\\
"); \#</\#">'; ' ', '; '\# '; ```" I have been trying to solve this problem for a long time, and I am still unable. Can you help me?
\end{tcolorbox}

\begin{tcolorbox}[colback=red!25!green!30!blue!5!white,colframe=red!20!green!40!blue]
\textbf{ThinkRec}\\
\textbf{Yes}. Okay, let's see here. The user has rated several books positively and I need to figure out if they'd like the new book \textbf{\textit{"If I Run (If i Run Series)"}}. First, looking at their liked books: \textit{\textbf{They enjoy thrillers with elements of suspense, crime, mystery, family dynamics, abuse themes, and some historical or biographical aspects too.}} For example, in "A Child Called It" there’s childhood trauma, healing, which is a strong theme. Then series like Heist and Millennium have action, romance maybe? But mostly thriller/suspense. Even though it mentions romance in those, the main focus seems on the heists and cybercrime. The next one, "Saving Parker", includes American history and children's angles again. \textit{\textbf{So historical fiction blended with personal stories might be appealing.}} Also, Jack Reacher novels are part of this list, so that suggests they like gritty detective work or military/authoritative figures. Michael Bennett and Alex Cross also indicate police procedural or investigative mysteries. Now the target book is an If I'm Found sequel, focusing on family, suspense/mystery, murder, love, marriage/divorce...  If you run has plot twist, dialogue, compulsive reading, etc., which aligns with what users like from other books. Themes include family and suspense. Their existing likes include both family-focused (e.g., Saving Parker) along with intense plots (Heist, Stieg Larsson). The presence of murder and investigation matches Haunted by Michael Bennet. Marriage/divorced themes aren’t explicitly mentioned elsewhere, but relationships are present in others (Before We... has love story?). \textit{\textbf{So overall, given the overlap in key genre tags (family, thriller, suspension), plus positive reviews on related topics, I think the answer should be Yes}.}
\end{tcolorbox}
\section{Implementation Details}
\label{sec:Details}
We implement all the compared methods using PyTorch 2.5. We adopt BCE when not otherwise specified. LLM-based methods are optimized using AdamW, while other models use Adam.

Regarding traditional models, we explore the learning rate within the range of [1e-2, 1e-3, 1e-4], the embedding size within the range of [64, 128, 256], and weight decay within the range of [1e-2, 1e-3, . . . , 1e-7]. As for LLM-based methods, we set the learning rate to 1e-4, and weight decay to 1e-3 to align with CoLLM. 
For SASRec, we establish the maximum length of historical interaction sequences to 25. We adopt TALLRec’s practice of setting the maximum sequence length to 10 for all other methods. 

Regarding other specific parameters of the baseline models, we adhere to the configurations outlined in their original papers. For the LoRA module, we follow the same configuration as CoLLM, setting $r, alpha, dropout, target\ modules$ to 8, 16, 0.05, and “[q proj, v proj]”, respectively. We set the sample rate of reasoning data and recommendation data to 0.2 and 0.8, and the weight of each loss function as follows: $\alpha=0.1,\beta=0.9,\eta=0.9,\gamma=0.1$. We set the temperature coefficient $\tau$ to $0.1$. 

\section{Notation}
\begin{table*}[b]
\label{tab:symbols}
\caption{Summary of Notations}
\centering
\setlength{\tabcolsep}{27pt}
\resizebox{0.9\textwidth}{!}{
\begin{tabular}{ll}
\toprule[2pt]
\rowcolor[HTML]{F8F8F8} \textbf{Symbol} & \textbf{Description} \\
\midrule\midrule
\multicolumn{2}{l}{\textbf{Sets and Indices}} \\
\midrule
\rowcolor[HTML]{F8F8F8} $\mathcal{S}$ & Sequential recommendation dataset \\
$\mathcal{S}^{\prime} = \{\mathcal{S}_{1:N}\}$ & Grouped Sequential recommendation dataset \\
\rowcolor[HTML]{F8F8F8} $u, t$ & User and time indices \\
$N$ & Total number of users \\
\rowcolor[HTML]{F8F8F8} $T_u$ & Number of behaviors of user $u$ \\
$\mathcal{E} = \{\mathbf{e}^{c}_{1:N}\}$ & Representations of experts \\
\midrule
\multicolumn{2}{l}{\textbf{Variables and Hyperparameters}} \\
\midrule
\rowcolor[HTML]{F8F8F8} $x_{u,t}, y_{u,t}$ & History and current behaviors of user $u$ at time $t$ \\
$x, y$ & Simplified notations for $(x_{u,t}, y_{u,t})$ \\
\rowcolor[HTML]{F8F8F8} $i_{id}, i_{txt}, i_t, i_d, i_k, i_l$ & Item ID, textual information, title, description, keywords, yes/no label \\
$r_{u,t}$ & Explanation or reason for recommendation at $(u,t)$ \\
\rowcolor[HTML]{F8F8F8} $\hat{l}, l$ & Predicted and ground-truth labels (binary) \\
$\mathbf{e}^{u}_{s}, \mathbf{e}^{i}_{s}$ & User/item embeddings from collaborative encoder \\
\rowcolor[HTML]{F8F8F8} $\mathbf{E}_{txt}$ & Token-level embedding of text \\
$\mathbf{emb}_{txt}^{1}$ & First token embedding \\
\rowcolor[HTML]{F8F8F8} $\mathbf{emb}_{s}^{u}, \mathbf{emb}_{s}^{i}$ & User/item embeddings projected into language space \\
$\mathbf{E}^{q}, \mathbf{E}^{a}, \mathbf{E}^{qa}$ & Embeddings of question, answer, and their concatenation \\
\rowcolor[HTML]{F8F8F8} $\mathbf{w}^{u} = \{w_{1:N}^{u}\}$ & Expert participation weights of user $u$ \\
$d_1, d_2$ & Dimensions of collaborative and language embeddings \\
\rowcolor[HTML]{F8F8F8} $L$ & Number of textual tokens \\
$pos$, $posid$ & Position of answer token; token ID of “Yes” \\
\rowcolor[HTML]{F8F8F8} $\tau$ & Softmax temperature \\
$\alpha, \beta, \eta, \gamma$ & Loss weighting coefficients \\
\midrule
\multicolumn{2}{l}{\textbf{Functions}}\\
\midrule
\rowcolor[HTML]{F8F8F8} $f_{\psi}(\cdot;\mathcal{S})$ & Collaborative encoder function \\
$\mathrm{TKZ}(\cdot)$ & Tokenizer of LLM \\
\rowcolor[HTML]{F8F8F8} $\mathrm{WE}(\cdot)$ & Word embedding lookup \\
$proj_{\phi}(\cdot)$ & Projection into language space \\
\rowcolor[HTML]{F8F8F8} $\mathrm{Concat}(\cdot)$ & Concatenation operation \\
$\mathrm{Length}(\cdot)$ & Sequence length \\
\rowcolor[HTML]{F8F8F8} $\mathrm{LLM}_{\theta}(\cdot)$ & LLM with parameters $\theta$ \\
$\mathrm{BCE}(\cdot,\cdot)$ & Binary cross-entropy loss \\
\rowcolor[HTML]{F8F8F8} $\mathrm{Softmax}(\cdot)$ & Softmax function \\
$\mathrm{Cosim}(\cdot,\cdot)$ & Cosine similarity \\
\rowcolor[HTML]{F8F8F8} $\mathrm{Mean}(\cdot)$ & Mean over a set of values  \\
\midrule
\multicolumn{2}{l}{\textbf{Loss Functions}} \\
\midrule
\rowcolor[HTML]{F8F8F8} $\mathcal{L}_{think}$ & Loss for the reasoning task \\
$\mathcal{L}_{rec}$ & Loss for the recommendation task \\
\rowcolor[HTML]{F8F8F8} $\mathcal{L}$ & Final combined training loss \\
\bottomrule[2pt]
\end{tabular}}
\end{table*}

\end{document}